\documentclass[preprint,nocite]{epl}
\usepackage{amsmath}

\title{Comment on ``Stochastic resonance in a linear system
with signal-modulated noise''}

\author{P.  F.  G\'ora\thanks{\email{gora@if.uj.edu.pl}}}

\institute{
M.~Smoluchowski Institute of Physics and Complex Systems 
Research Center, Jagellonian University, Reymonta~4, 30-059~Krak\'ow, Poland
}

\begin{document}

\maketitle

\begin{abstract}
The system discussed recently in
L.~Cao and D.~J.~Wu, Europhys. Lett. {\bf 61}, (2003) 593 
merely reproduces properties of the input signal.
\end{abstract}

Stochastic resonance (SR) is an example of a~constructive role of noise, 
where the noise and a
dynamical system act together to reinforce the output of the latter.
Recently, Cao and Wu in Ref.~\cite{cao} discussed an overdamped, linear system 
with purely
additive, correlated, coloured noises, partially modulated by a periodic signal, 
and observed a phenomenon which they
described as SR. This effect indeed fits into a broad definition of SR that
some parameters of the system are optimized by a specific level of the stochastic
forcing \cite{mcclintock,moss} but the authors of Ref.~\cite{cao} do not 
provide any qualitative interpretation of their findings. The dynamical system
is governed by the following equation:

\begin{equation}\label{epl:system}
\dot x = -\alpha x + f(t) = -\alpha x + A_0 \xi(t) \cos(\Omega t+\phi) + \zeta(t)\,,
\end{equation}

\noindent where $\zeta(t) = \xi(t) + \eta(t)$ and 
$\left\langle\xi(t)\right\rangle = \left\langle\eta(t)\right\rangle = 0$,
$\left\langle\xi(t)\xi(t^\prime)\right\rangle = Q u(|t-t^\prime|)$,
$\left\langle\eta(t)\eta(t^\prime)\right\rangle = D u(|t-t^\prime|)$,
$\left\langle\xi(t)\eta(t^\prime)\right\rangle = 2\lambda\sqrt{QD} u(|t-t^\prime|)$.
In Ref.~\cite{cao} $u(|t-t^\prime|) = \exp(-|t-t^\prime|/\tau)/\tau$, 
but as we shall see, the choice of $u$ is
not particularly important. Cao and Wu calculate 
$\left\langle x(t)x(t+t^\prime)\right\rangle$ (but do not provide the
final formula for it), use the Wiener-Khinchin
theorem to calculate the power spectrum, and conclude that SR
is present in the system.

While the authors of Ref.~\cite{cao} do not state that explicitly, we assume
that the correlation function has been averaged over the initial phase, $\phi$,
as otherwise it would produce a time-dependent spectrum whose interpretation is
not obvious \cite{Honerkamp,Wiesenfeld}. These authors also use a rather non-standard measure
of the SR. But the crucial observation is that
regardless of whether we admit time-dependent power spectra
or not and what definition 
of SR we use, the signal-to-noise ratio (SNR) for the 
system \eqref{epl:system} with \textit{any} 
correlation function $u(|t-t^\prime|)$ has the form

\begin{equation}\label{epl:SNR}
\text{SNR} = \frac{p\,Q}{Q+D+2\lambda\sqrt{QD}},
\end{equation}

\noindent where the coefficient $p$ may depend on time, the characteristic frequency
$\Omega$ and other factors, cf.\ Eqs.\ (14)-(16) in Ref.~\cite{cao}.
If we now calculate the correlation function
of the input in \eqref{epl:system} and average it over 
the initial phase $\phi$

\begin{equation}\label{epl:inputcorrelation}
\frac{1}{2\pi}\int\limits_0^{2\pi}\left\langle f(t)f(t+t^\prime)\right\rangle d\phi
= \frac{1}{2}A_0^2Qu(|t^\prime|) \cos\Omega t^\prime + 
(Q+D+2\lambda\sqrt{QD}) u(|t^\prime|)\,,
\end{equation}

\noindent we get for the SNR
\textit{of the input signal} an expression of precisely the same form as 
\eqref{epl:SNR} but with $p$ replaced by some other coefficient $\tilde p$. 
The same result may be obtained directly from \eqref{epl:inputcorrelation}
by comparing the amplitudes of the oscillatory term and the noisy background.
Since there is no feedback, features of the input signal are oblivious to 
any dynamics that the system \eqref{epl:system} may display. 
Any resonant behaviour corresponding to changes in $Q$ can be present in the 
output if and only if it is present in the input. Thus, the resonant behaviour 
of the output solely reflects changes in spectral properties of the input signal.

For $\lambda<0$ the function \eqref{epl:SNR} displays a maximum and even
diverges for $\lambda=-1$ and $Q=D$, which is unphysical.
The reason for this behaviour of SNR is that for negative $\lambda$ the 
variance of the unmodulated noise $\zeta(t)$ may decrease as the variance of 
the stochastic amplitude of the periodic signal increases --- observe that $Q$ 
enters the variances of both the noises. The fact that one 
cannot independently manipulate various noise strengths is unwelcome
and we should think of a better parameterisation of the problem.
If instead of \eqref{epl:system} we write

\begin{equation}\label{epl:system1}
\dot x=-\alpha x + \sqrt{Q}\,\xi(t)\cos(\Omega t+\phi) 
+ \sqrt{D}\,\zeta(t)
\end{equation}

\noindent where $\zeta(t) = \lambda\,\xi(t) +\sqrt{1-\lambda^2}\,\eta(t)$,
$\lambda\in[-1,1]$ and

\begin{subequations}\label{epl:noises1}
\begin{gather}
\left\langle\xi(t)\right\rangle = \left\langle\eta(t)\right\rangle = 
\left\langle\xi(t)\eta(t^\prime)\right\rangle =0\,,\\
\left\langle\xi(t)\xi(t^\prime)\right\rangle =
\left\langle\eta(t)\eta(t^\prime)\right\rangle = u(|t-t^\prime|)\,,
\end{gather}
\end{subequations}

\noindent we capture the most essential features of the system. Now
the variances of the unmodulated and signal-modulated noises are independent, 
we do not risk any divergences, but we do not have
any resonant behaviour in the input signal, either.

In conclusion, we have shown that the dynamics of the system described in
Ref.~\cite{cao} merely reproduces properties of the
input signal. Moreover, the resonant properties of the input signal in are
an artifact of a not particularly successful parameterisation of the noise.

\end{document}